\documentclass[preprintnumbers,aps,prb,twocolumn,showpacs]{revtex4}

\usepackage{graphicx}
\usepackage{dcolumn} 
\usepackage{amsmath}
\usepackage{exscale}
\usepackage{bm}      
\usepackage{color}
\usepackage{epsfig,psfrag,subfigure,subeqnarray}
\usepackage{bbm}
\usepackage{amsbsy}
\usepackage{amssymb}
\usepackage{enumitem}

\def\lan{\langle}
\def\ran{\rangle}
\def\va{\varepsilon}

\def\vk{{\bf k}}
\def\vK{{\bf K}}

\def\vq{{\bf q}}
\def\vp{{\bf p}}

\newcommand{\bd}{\begin{equation}}
\newcommand{\ed}{\end{equation}}
\newcommand{\be}{\begin{equation}}
\newcommand{\ee}{\end{equation}}
\newcommand{\bt}{\begin{split}}
\newcommand{\et}{\end{split}}

\newcommand{\bn}{\begin{align}}
\newcommand{\en}{\end{align}}

\newcommand{\bea}{\begin{eqnarray}}
\newcommand{\eea}{\end{eqnarray}}
\newcommand{\ba}{\begin{array}}
\newcommand{\ea}{\end{array}}
\newcommand{\nn}{\nonumber}

\begin{document}
\title {Coboson many-body formalism for cold atom dimers with attraction between different fermion species only}

\author{ Monique Combescot$^1$, Shiue-Yuan Shiau$^{2}$ and Yia-Chung Chang$^{3,2}$}
\email{yiachang@gate.sinica.edu.tw}
\affiliation{$^1$Institut des NanoSciences de Paris, Universit\'e Pierre et Marie Curie, CNRS, 4 place Jussieu, 75005 Paris}

\affiliation{$^2$ Department of Physics, National Cheng Kung University, Tainan, 701 Taiwan}
\affiliation{$^3$Research Center for Applied Sciences, Academia Sinica, Taipei, 115 Taiwan}

\date{\today}

\begin{abstract}
Unlike the Coulomb potential that acts between all semiconductor carriers, the potential commonly used for BCS superconductors and cold atom gases acts between different fermion species only, these species differing by their spin or hyperfine index. The coboson formalism we here develop evidences that such composite bosons interact through fermion exchange only, thus rendering the structure of the Shiva diagrams that visualize their many-body effects far simpler. A separable potential is used to obtain analytical results for the $N$-coboson normalization factor and the $N$-coboson ground-state energy within the Born approximation, in terms of the Born dimer-dimer scattering length. This formalism opens the route toward approaching complex many-body effects, such as Bose-Einstein condensation, through a new perspective.

\end{abstract}


\maketitle

\section{Introduction\label{sec:1}}

In the early 2000's, a many-body formalism\cite{M_PR2008} has been developed to handle composite bosons (cobosons) as entities, while keeping fermion exchange between these particles in an exact way. This formalism has been applied to predict various physical effects involving excitons that occur in semiconductor nonlinear optics\cite{M_SSC2004,M_EP2005,M_PRB2006}.    \

To fully take into account fermion exchanges between the fermionic components of composite particles may seem at first quite daunting. This could be the reason why the many-body formalism for elementary fermions, developed half a century ago, is still used for systems made of composite quantum particles. It turns out that the many-body formalism for composite particles that has been proposed by one of the authors is not all that complicated. Instead of a scalar algebra based on Green functions, it uses an operator algebra based on commutators in the case of composite bosons and commutators and anticommutators in the case of composite fermions \cite{M_PRL2010,Marxiv}. These commutation relations involve the system Hamiltonian of the constituting fermions and the creation operators for single composite particles. Shiva diagrams have been proposed to visualize many-body effects occurring between composite bosons made of two fermions, while more complicated Kali diagrams are necessary for composite fermions. Like Feynman diagrams, these diagrams are powerful in the sense that they allow us to directly write down the physical quantities they represent.\

The coboson formalism has so far been mostly used for problems concerning semiconductor excitons, with long-range Coulomb attraction between electrons and holes as large as Coulomb repulsions between electrons and between holes. In other fields like BCS superconductors and cold atom gases, however, the potential is short-range, and commonly approximated as an attraction between different fermion species only. We here present the coboson many-body formalism appropriate for such systems. The reduction of their potential to an attraction between different fermion species greatly simplifies the coboson many-body formalism because this potential in fact corresponds to a one-body operator in the coboson subspace. Such a potential, which also is the one used for Cooper pairs in BCS superconductivity\cite{M_EPJB2011}, leads to a many-body physics driven by fermion exchanges, thus making the associated Shiva diagrams far simpler. In view of the physical importance of the fields in which the potential can be restricted to that between different fermions only, it is of interest to present this simple version of the coboson many-body formalism.    \

The paper is organized as follows:

In Sec.~\ref{sec:2}, we present the coboson many-body formalism for a two-fermion system having an attractive potential between different fermion species.\

In Sec.~\ref{sec:3}, we take the potential scattering in a separable form in order to derive explicit results. Such a scattering is commonly taken for short-range attraction such as the one acting between up-spin and down-spin electrons making Cooper pairs, or the one used for cold atom gases. Within such a separable potential, we calculate two basic quantities, namely the normalization factor for $N$ ground-state cobosons and the Hamiltonian mean value in this state. The normalization factor provides a direct way to gauge the ``entanglement" existing between composite bosons\cite{CKL2005,Chudzicki2010,Tichy2012,M_EPL2011}. Its reduction with increasing $N$ originates from the ``moth-eaten effect" induced by the Pauli exclusion principle between fermionic components\cite{PogosovJETP2010}. The Hamiltonian mean value corresponds to the system ground-state energy in the Born approximation. We find that it has higher than linear density dependence, in contrast to that of Cooper pairs\cite{M_EPJB2011_2,Crouzeix2011}, in which higher density terms cancel exactly because in addition to being present between opposite spins only, the attraction occurs only between pairs having zero center-of-mass momentum. We find that the linear term in density of the Hamiltonian mean value corresponds to the well-known Born term, $2a$, for the scattering length between dimers obtained from standard Green function procedure\cite{Haussmann1993,Melo1993} and partial-bosonization procedure\cite{PieriPRB2000}.\

In the last section, we discuss the advantage of the coboson formalism and conclude.\

\section{General coboson formalism\label{sec:2}}
We consider a system made of two fermion species, $\alpha$ and $\beta$, with an attractive force between $(\alpha, \beta)$ fermions only. The system Hamiltonian, $H=H_0+V$, contains a free part given by
\be
H_0=\sum_{\vk_\alpha} \va^{(\alpha)}_{\vk_\alpha} a^\dag_{\vk_\alpha}a_{\vk_\alpha}+\sum_{\vk_\beta} \va^{(\beta)}_{\vk_\beta} b^\dag_{\vk_\beta}b_{\vk_\beta}\label{eq:H0}
\ee
with $\va^{(\alpha,\beta)}_\vk=\vk^2/2m_{\alpha,\beta} $, and an interaction part given by
\be
V=-\sum_\vq v_\vq \sum_{\vk_\alpha \vk_\beta} a^\dag_{\vk_\alpha+\vq}b^\dag_{\vk_\beta-\vq}b_{\vk_\beta} a_{\vk_\alpha}\,\label{eq:V0} .
\ee
(Throughout this paper, $\hbar$ is set equal to 1.)

\subsection{Single-pair operators}

The coboson many-body formalism is based on the creation operators for single pair eigenstates, $|i\ran=B^\dag_i|v\ran$,
\be
0=(H-E_i)|i\ran\, .
\ee
Just like free pair states $|\vk_\alpha,\vk_\beta\ran$, which are eigenstates of the free Hamiltonian $H_0$,
\be
0=(H_0-\va^{(\alpha)}_{\vk_\alpha}-\va^{(\beta)}_{\vk_\beta})|\vk_\alpha,\vk_\beta\ran\, ,
\ee
the correlated pair states $|i\ran$, which are eigenstates of the interacting Hamiltonian $H$, form a complete basis in the one-pair subspace
\be
I_1=\sum_i|i\ran\lan i| = \sum_{\vk_\alpha \vk_\beta}|\vk_\alpha,\vk_\beta\ran \lan\vk_\beta,\vk_\alpha |\, .
\ee
So, correlated pair and free pair states are related by
\bea
|i\ran&=& \sum_{\vk_\alpha \vk_\beta}|\vk_\alpha,\vk_\beta\ran \lan\vk_\beta,\vk_\alpha |i\ran\, ,\label{eq:itokakb} \\
 |\vk_\alpha\vk_\beta\ran &=& \sum_i |i\ran\lan i|\vk_\alpha,\vk_\beta\ran\, .\label{eq:kakbtoi}
\eea

As total momentum is conserved in the interaction processes of any potential for translationally invariant systems, the relevant fermion pair operator must have the center-of-mass momenta $\vK$ appearing explicitly, as in
\be
B^\dag_{\vK\vp}=a^\dag_{\vp+\gamma_\alpha \vK}b^\dag_{-\vp+\gamma_\beta \vK}\label{eq:defBKp}
\ee
with $\gamma_\alpha=1-\gamma_\beta= m_\alpha/(m_\alpha+m_\beta)$ and $\vp$ being the relative motion momentum of the pair. The pair kinetic energy then reduces to
\be
\va^{(\alpha)}_{\vp+\gamma_\alpha \vK}+\va^{(\beta)}_{-\vp+\gamma_\beta \vK}=\frac{\vK^2}{2M}+\frac{\vp^2}{2\mu}
\ee
with $M=m_\alpha+m_\beta$ and $\mu^{-1}=m^{-1}_\alpha+m^{-1}_\beta $. By writing $\vk_\alpha=\vp+\gamma_\alpha \vK$ and $\vk_\beta=-\vp+\gamma_\beta \vK$ in Eq.~(\ref{eq:V0}), we can rewrite this equation as
\be
V=-\sum_\vq v_\vq \sum_{\vK \vp} B^\dag_{\vK, \vp+\vq} B_{\vK \vp}=-\sum_{\vK\vp\vp'} B^\dag_{\vK\vp'} v_{\vp'-\vp}B_{\vK \vp}\, .\label{eq:VintofBKp}
\ee

Since the potential $V$ conserves momentum, the correlated pair eigenstates are a linear combination of free pairs having same center-of-mass momentum. So,
\be
|i\ran\equiv |\vK_i,\nu_i\ran=\sum_{\vp}|\vK_i \vp\ran\lan \vp|\nu_i\ran
\ee
with $|\vK_i \vp\ran=B^\dag_{\vK_i \vp}|v\ran$, where $B^\dag_{\vK \vp}$ is the operator defined in Eq.~(\ref{eq:defBKp}). For $V$ given in Eq.~(\ref{eq:VintofBKp}), the relative motion wave function $\lan \vp|\nu_i\ran$ of the coboson $i$ eigenstate of $H$,
\be
0=(H_0+V-E_i)\sum_{\vp}|\vK_i \vp\ran\lan \vp|\nu_i\ran\, ,
\ee
 must fulfill
\be
0=(\va_{\vp'}-\va_{\nu_i})\lan \vp'|\nu_i\ran-\sum_\vp v_{\vp'-\vp}\lan \vp|\nu_i\ran \label{eq:schroeq_1dimer}
\ee
with $E_i=\va_{\nu_i}+\vK_i^2/2M$ and $\va_\vp=\vp^2/2\mu$.\

\begin{figure}[b]
\centering

   \includegraphics[trim=5cm 6.7cm 5cm 20.2cm,clip,width=3in] {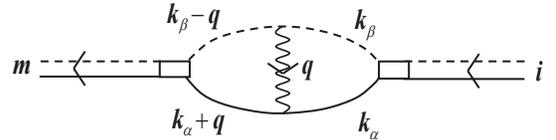}
 \vspace{-0.1cm}
\caption{\small Scattering amplitude $v_{mi}$ resulting from interaction between fermions $(\alpha,\beta)$, as given by Eq.~(\ref{eq:vmi}). Fermions $\alpha$ are represented by solid lines and fermions $\beta$ by dashed lines. To ``open" the coboson $i$ into free pairs, we pay $\lan\vk_\beta,\vk_\alpha |i\ran$. These free pairs then interact. To close the resulting free pairs into a coboson $m$, we pay $\lan m|\vk_\alpha+\vq, \vk_\beta-\vq\ran$, which is what Eq.~(\ref{eq:vmi}) tells. }
\label{fig:vmi_scatt}
\end{figure}

Equation (\ref{eq:VintofBKp}) also leads, using Eq.~(\ref{eq:kakbtoi}), to
\bea
V|i\ran&=& -\sum_\vq v_\vq \sum_{\vk_\alpha \vk_\beta} |\vk_\alpha+\vq,\vk_\beta-\vq\ran \lan\vk_\beta,\vk_\alpha |i\ran\nn\\
&=& -\sum_m |m\ran v_{mi}\, ,
\eea
where the scattering between correlated-pair states is given by
\be
v_{mi}= \sum_\vq v_\vq  \sum_{\vk_\alpha \vk_\beta}\lan m|\vk_\alpha+\vq, \vk_\beta-\vq\ran \lan\vk_\beta,\vk_\alpha |i\ran\, .\label{eq:vmi}
\ee
This scattering is visualized by the diagram of Fig.~\ref{fig:vmi_scatt}.

We now turn to pair operators on which the coboson many-body formalism is constructed. Equations (\ref{eq:itokakb}) and (\ref{eq:kakbtoi}) allow us to relate correlated pair operators to free pair operators through
\bea
B^\dag_i&=& \sum a^\dag_{\vk_\alpha} b^\dag_{\vk_\beta} \lan\vk_\beta,\vk_\alpha  |i\ran\,\label{eq:Btoab} ,\\
a^\dag_{\vk_\alpha}b^\dag_{\vk_\beta}&=& \sum B^\dag_i \lan i|\vk_\alpha,\vk_\beta \ran \, .\label{eq:abtoB}
\eea
By writing free pair operators in terms of correlated pair operators, the potential acting only between $(\alpha, \beta)$ fermions reduces
\be
V=-\sum_{mi}v_{mi}B^\dag_m B_i\, ,\label{eq:V}
\ee
where $v_{mi}$ is the scattering amplitude given in Eq.~(\ref{eq:vmi}). The above equation demonstrates that the potential between different fermions $(\alpha, \beta)$ is a one-body operator in the coboson subspace. As a direct consequence, interaction between cobosons can only come from fermion exchanges induced by the Pauli exclusion principle.

\subsection{Pauli scatterings}
\begin{figure}[h]
\centering
   \includegraphics[trim=6.2cm 6.7cm 6.4cm 19.1cm,clip,width=2.6in] {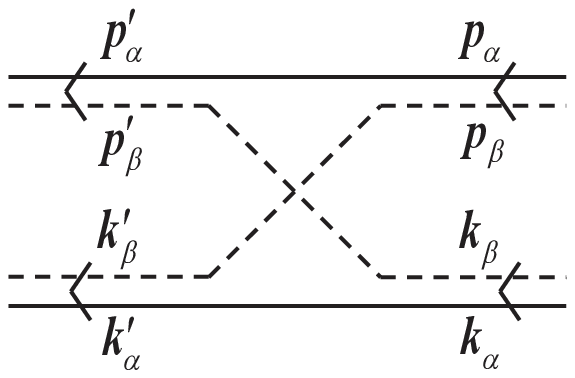}
	\vspace{-0.2cm}
\caption{\small Pauli scattering for fermion exchange between free pairs, $\lambda\big(^{\hspace{.06cm}
\vp'_\alpha \vp'_\beta \hspace{.12cm} \vp_\alpha \vp_\beta \hspace{.05cm}}_{\hspace{.06cm}
\vk'_\alpha \vk'_\beta \hspace{.12cm} \vk_\alpha \vk_\beta\hspace{.05cm}}\big)$, as given in Eq.~(\ref{eq:lambdakk}).  }
\label{fig:lambdakk}
\end{figure}

To derive the Pauli scatterings between two cobosons resulting from fermion exchanges between correlated pairs in the absence of fermion-fermion interaction, it is convenient to start with free pairs with creation operators $a^\dag_{\vk_\alpha} b^\dag_{\vk_\beta}=\cal{B}^\dag_{\vk_\alpha \vk_\beta}$. We first construct their ``deviation operators", $D$, through
\be
\Big[\cal{B}_{\vk'_\alpha \vk'_\beta},\cal{B}^\dag_{\vk_\alpha \vk_\beta}  \Big]_-=\delta_{\vk'_\alpha \vk_\alpha}\delta_{\vk'_\beta \vk_\beta}-D_{\vk'_\alpha \vk'_\beta,\vk_\alpha \vk_\beta}\, ,
\ee
from which we get
\bea
\lefteqn{\Big[D_{\vk'_\alpha \vk'_\beta,\vk_\alpha \vk_\beta},\cal{B}^\dag_{\vp_\alpha \vp_\beta}  \Big]_-=}\label{eq:DB_commu}\\
&&\sum_{\vp'_\alpha \vp'_\beta} \cal{B}^\dag_{\vp'_\alpha \vp'_\beta}\bigg( \lambda\left(\begin{smallmatrix}
\vp'_\alpha \vp'_\beta & \vp_\alpha \vp_\beta \\ \vk'_\alpha \vk'_\beta & \vk_\alpha \vk_\beta\end{smallmatrix}\right)+\lambda\left(\begin{smallmatrix}
\vp'_\alpha \vp'_\beta &\vk_\alpha \vk_\beta  \\ \vk'_\alpha \vk'_\beta & \vp_\alpha \vp_\beta\end{smallmatrix}\right) \bigg)\, .\nn
\eea
The Pauli scattering between two free pairs, as visualized by the diagram of Fig.~\ref{fig:lambdakk}, reduces to momentum conservation,
\be
 \lambda\left(\begin{smallmatrix}
\vp'_\alpha \vp'_\beta & \vp_\alpha \vp_\beta \\ \vk'_\alpha \vk'_\beta & \vk_\alpha \vk_\beta\end{smallmatrix}\right)=\delta_{\vk'_\alpha \vk_\alpha} \delta_{\vk'_\beta \vp_\beta  }\delta_{\vp'_\beta \vk_\beta } \delta_{\vp'_\alpha \vp_\alpha}\, ,\label{eq:lambdakk}
\ee
 as can be directly read from the figure.
The two $\lambda$ terms in Eq.~(\ref{eq:DB_commu}) correspond to an exchange of $\alpha$ fermions and $\beta$ fermions, as expected from $(\alpha, \beta)$ symmetry.\

\begin{figure}[h]
\begin{center}
\includegraphics[trim=5.7cm 6.6cm 6.1cm 19.2cm,clip,width=2.6in] {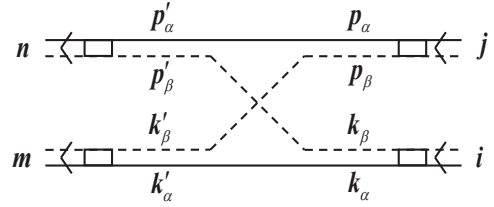}
\end{center}
\vspace{-0.6cm}
\caption{\small Pauli scattering $\lambda\big(^{\hspace{.06cm}
n \hspace{.12cm} j\hspace{.05cm}}_{\hspace{.05cm}
m\hspace{.09cm} i\hspace{.05cm}}\big)$ for fermion exchange between correlated pairs, as given in Eq.~(\ref{def_lambda}). To get it, we ``open" the correlated pairs into free pairs through $\langle\vk_\beta, \vk_\alpha |i\rangle$ factors, let the resulting free pairs exchange their fermions, and then ``close" the free pairs into correlated pairs through $\lan m| \vk_\alpha', \vk_\beta'\ran$.}
\label{fig:lambda_mnij}
\end{figure}

The link between correlated and free pair operators given in Eqs.~(\ref{eq:Btoab}) and (\ref{eq:abtoB}) then readily gives, for normalized correlated pair states $\lan m|i\ran=\delta_{mi}$,
\bea
\Big[B_m,B^\dag_i\Big]_-&=& \delta_{mi}-D_{mi}\, ,\label{eq:commut_BB}\\
D_{mi}&=& \sum_{\{\vk\}} \lan m| \vk'_\alpha ,\vk'_\beta\ran D_{ \vk'_\alpha \vk'_\beta,  \vk_\alpha \vk_\beta} \lan \vk_\beta, \vk_\alpha |m\ran\, ,\hspace{0.6cm}
\eea
from which we get the Pauli scatterings between correlated pairs as
\bea
\left[D_{mi},B^\dag_j\right]_-&=&\sum_n\Big(\lambda\left(\begin{smallmatrix}
n& j\\ m& i\end{smallmatrix}\right)+\lambda\left(\begin{smallmatrix}
n& i\\ m& j\end{smallmatrix}\right)\Big)B_n^\dag\, ,\label{eq:commut_DB}\\
\lambda\left(\begin{smallmatrix}
n& j\\
m& i
\end{smallmatrix}\right)&=&\sum\langle m|\vk'_\alpha,\vk'_\beta\rangle\langle n|\vp'_\alpha,\vp'_\beta\rangle \lambda\left(\begin{smallmatrix} \vp'_\alpha \vp'_\beta & \vp_\alpha \vp_\beta \\ \vk'_\alpha \vk'_\beta & \vk_\alpha \vk_\beta\end{smallmatrix}\right)\nn\\
&& \times \langle\vp_\beta, \vp_\alpha |j\rangle\langle\vk_\beta, \vk_\alpha |i\rangle. \label{def_lambda}
\eea
The above expression of $\lambda\left(\begin{smallmatrix} n& j\\ m& i \end{smallmatrix}\right)$ can be directly read from the diagram of Fig.~\ref{fig:lambda_mnij}  which represents the Pauli scattering induced by fermion exchange between cobosons $(i,j)$ that end in states $(m,n)$.

\subsection{Interaction Scatterings}

We now turn to scatterings between two cobosons resulting from the potential $V$. From Eqs.~(\ref{eq:H0}) and (\ref{eq:Btoab}), we get
\be
\left[H_0,B^\dag_i\right]_-=\sum \left(\va^{(\alpha)}_{\vk_\alpha}+\va^{(\beta)}_{\vk_\beta}\right)B^\dag_{\vk_\alpha \vk_\beta} \langle\vk_\beta ,\vk_\alpha |i\rangle\, ,
\ee
while using Eq.~(\ref{eq:commut_BB}) and the form of $V$ given in Eq.~(\ref{eq:V}), we readily get
\be
\left[V,B^\dag_i\right]_-=-\sum_m v_{mi} B^\dag_m +\sum_{mm'} B^\dag_m v_{mm'} D_{m' i}\, .
\ee
By noting that
\bea
0&=& \lan\vk_\beta,\vk_\alpha|H_0+V-E_i |i\ran\nn\\
&=& (\va^{(\alpha)}_{\vk_\alpha}+\va^{(\beta)}_{\vk_\beta}-E_i)  \lan\vk_\beta,\vk_\alpha  |i\ran-\sum_m  \lan\vk_\beta,\vk_\alpha  |m\ran v_{mi}\, ,\hspace{0.6cm}\label{eq:dimerSchroequ}
\eea
we obtain
\be
\left[H,B^\dag_i\right]_-=E_iB^\dag_i +V^\dag_i\,\label{eq:HB_commu} ,\ee
where the creation potential for coboson $i$ is given by
\be
V^\dag_i=  \sum_{mm'} B^\dag_m v_{mm'} D_{m' i}\, .
\ee

\begin{figure}[h]
\begin{center}
\includegraphics[trim=3.3cm 6.9cm 2.4cm 19.2cm,clip,width=3.5in] {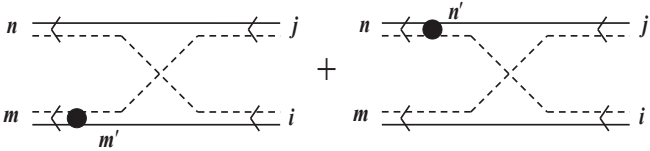}
\end{center}
\vspace{-0.5cm}
\caption{{\small The first two terms of the interaction scattering $\xi\big(^{\hspace{.06cm}
n \hspace{.12cm} j\hspace{.05cm}}_{\hspace{.05cm}
m\hspace{.09cm} i\hspace{.05cm}}\big)$ given in Eq.~(\ref{eq:xi_mnij}). The other two terms come from exchanging the $\alpha$ fermions, instead of the $\beta$ fermions. This interaction scattering involves one interaction scattering but results from fermion exchange only. }}
\label{fig:Xi_mnij}
\end{figure}

The interaction scattering between cobosons in states $i$ and $j$ follows from one more commutator. Using Eq.~(\ref{eq:commut_DB}), we find
\bea
\left[V^\dag_i,B^\dag_j\right]_-&=& \sum_{mm'} B^\dag_m v_{mm'} \left[D_{m' i}, B^\dag_j\right]\nn\\
&=& \sum_{mm'n}B^\dag_m B^\dag_n v_{mm' } \Big(\lambda\left(\begin{smallmatrix} n& j\\ m'& i \end{smallmatrix}\right)+\lambda\left(\begin{smallmatrix} n& i\\ m'& j \end{smallmatrix}\right)\Big) \, .  \hspace{0.6cm}
\eea
It will be convenient to symmetrize the above equation with respect to $(m,n)$ in order to have
\be
\left[V^\dag_i,B^\dag_j\right]_-= \sum_{mn} B^\dag_m B^\dag_n \xi\left(\begin{smallmatrix} n& j\\ m& i \end{smallmatrix}\right)\label{eq:ViBjxi}
\ee
with $\xi\left(\begin{smallmatrix} n& j\\ m& i \end{smallmatrix}\right)=\xi\left(\begin{smallmatrix} m& j\\ n& i \end{smallmatrix}\right)$; the interaction scattering then reads
\bea
\xi\left(\begin{smallmatrix} n& j\\ m& i \end{smallmatrix}\right)&=&\frac{1}{2}\Big\{ \sum_{m'} v_{mm'}\lambda \left(\begin{smallmatrix} n& j\\ m'& i \end{smallmatrix}\right)+\sum_{n'} v_{nn'}\lambda \left(\begin{smallmatrix} n'& j\\ m& i \end{smallmatrix}\right) \nn\\
&&+ (i\longleftrightarrow j)\Big\}\, .\label{eq:xi_mnij}
\eea
As obvious from its Shiva diagram representation shown in Fig.~\ref{fig:Xi_mnij}, this interaction scattering evidences that cobosons $i$ and $j$ interact not because of the fermion-fermion potential, but because of fermion exchanges.
This peculiar result can be traced back to the fact that the potential $V$ acts between different fermion species only. Interaction scatterings coming from fermion exchange only also exist for Cooper pairs that interact via the reduced BCS potential: indeed, it has been shown\cite{PogosovJETP2010} that the many-body physics of Cooper pairs only comes from fermion exchanges, a point not often noted.\

\begin{figure}[h]
\begin{center}
\includegraphics[trim=2.3cm 4cm 2cm 17.2cm,clip,width=3.5in] {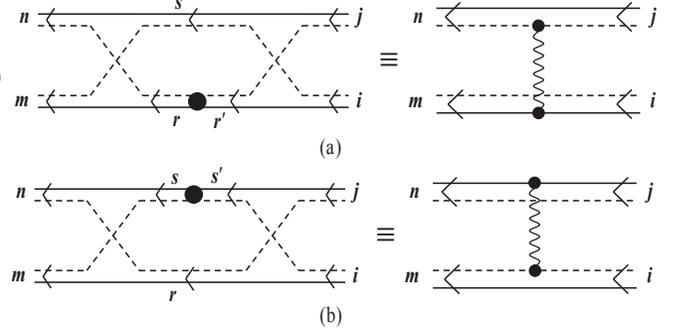}
\end{center}
\vspace{-0.5cm}
\caption{{\small (a,b) The two interaction scatterings contained in $\xi^{in}\big(^{\hspace{.06cm}
n \hspace{.12cm} j\hspace{.05cm}}_{\hspace{.05cm}
m\hspace{.09cm} i\hspace{.05cm}}\big)$ and defined in Eq.~(\ref{eq:xiin_mnij}). This scattering in fact corresponds to direct processes, the outgoing coboson $m$ being formed on the same fermion pair as the incoming coboson $i$. }}
\label{fig:Xiin_mnij}
\end{figure}

Another interaction scattering appearing in coboson many-body effects follows from the indistinguishability of the fermions that form cobosons. By adding one fermion exchange between cobosons to the above interaction scattering, we get
\be
\sum_{rs} \lambda\big(\begin{smallmatrix} n& s\\ m& r \end{smallmatrix}\big)\xi\big(\begin{smallmatrix} s& j\\ r& i \end{smallmatrix}\big)\equiv\xi^{in}\big(\begin{smallmatrix} n& j\\ m& i \end{smallmatrix}\big)\, ,\label{eq:xiin_mnij}
\ee
according to the standard notations of the coboson formalism\cite{M_PR2008}. In this scattering shown in Fig.~(\ref{fig:Xiin_mnij}), the two fermions which form coboson $i$ are the ones that form cobosons $m$, the $\alpha$ fermion of coboson $i$ interacting with the $\beta$ fermion of coboson $j$, with $(\alpha,\beta)$ possibly interchanged. \

To go further and obtain some analytical results, we must specify the potential acting between $(\alpha, \beta)$ fermions. As physically relevant potentials for superconductors and for cold atom gases are short-range, a convenient approximate form for such potentials is the so-called ``separable potential". We are going to use it in the following.

\section{Separable potential\label{sec:3}}
We first wish to note that the basic idea of a ``separable potential" is to separate incoming states from outgoing states in interaction processes that would otherwise be difficult to handle.
Taking a separable form for this potential then amounts to replacing $v_{\vp'-\vp}$ in Eq.~(\ref{eq:schroeq_1dimer}) with $v_0 w_{\vp'}w_{\vp}$ and taking $w_\vp$ nonzero for particle energy $\va_\vp$ in a designated range.

\subsection{Single-pair states}

{\bf $\bullet$ Bound state}\

\mbox{}

Let us first look for the existence of a bound state, i.e., a state below all possible free pair energies, which makes its energy $\va_{\nu_i}$ negative. For $v_{\vp'-\vp}=v_0 w_{\vp'}w_{\vp}$, Eq.~(\ref{eq:schroeq_1dimer}) leads to
\be
\lan \vp'|\nu_i\ran=\frac{w_{\vp'}}{\va_{\vp'}-\va_{\nu_i}} v_0 \sum_{\vp} w_{\vp} \lan \vp|\nu_i\ran\, .\label{eq:vpinui}
\ee
Multiplying the above equation by $w_{\vp'}$ and taking the sum over $\vp'$ gives the equation fulfilled by the single pair eigenvalues as
\be
\frac{1}{v_0}=\sum_{\vp'}\frac{w^2_{\vp'}}{\va_{\vp'}-\va_{\nu_i}}\, .\label{eq:singlepaireq}
\ee
For $|\va_{\nu_i}|$ large enough compared to the $\va_{\vp'}$ energy spacing, we can replace the discrete sum over $\vp'$ by an integral. The above equation has only one negative solution, i.e., one bound state. This bound state is the single-pair ground state $\nu_0$.\

Let us consider three standard model potentials:\\
(i) $w_\vp=1$ for $\va_{F_0}\leq \va_\vp \leq \va_{F_0}+\Omega$\

This corresponds to the potential used in BCS superconductivity. For a large Fermi energy $\va_{F_0}$ and a small potential layer extension $\Omega$, the density of states in the potential layer can be approximated by a constant, $\rho(\va)=\rho_0$. Equation (\ref{eq:singlepaireq}) then reduces to
\be
\frac{1}{v_0}=\int^{\va_{F_0}+\Omega}_{\va_{F_0}} \rho_0 \frac{d\va_\vp}{\va_\vp-\va_{\nu_0}}=\rho_0 \ln \frac{\va_{F_0}+\Omega-\va_{\nu_0}}{\va_{F_0}-\va_{\nu_0}}\,,
\ee
from which we find the single bound state solution as
\be
\va_{\nu_0}=\va_{F_0}-\frac{\Omega \sigma}{1-\sigma}\label{eq:sol:Cooperpair}
\ee
with $\sigma= e^{-1/\rho_0 v_0}$, as obtained by Cooper. (The factor of $2$ difference with respect to the standard exponent in $\sigma$ is due to the fact that $\rho_0$ here is the \emph{pair} density of states.)\\
(ii) $w_\vp=1$ for $0\leq \va_\vp \leq \Omega$ in 2D\

The 2D density of states being constant for whatever $\va_\vp$, the bound-state energy just follows from setting $\va_{F_0}=0$ in the above solution.\\
(iii)  $w_\vp=1$ for $0\leq \va_\vp \leq \Omega$ in 3D\

By writing the density of states as $\rho_0 \sqrt{\va_\vp /\Omega}$, the $\vp$ sum in Eq.~(\ref{eq:singlepaireq}), when transformed into an integral, leads to
\be
\frac{1}{2\rho_0 v_0} =1-\sqrt{\frac{-\va_{\nu_0}}{\Omega}}\tan^{-1} \sqrt{\frac{\Omega}{-\va_{\nu_0}}}\, .\label{eq:bounds_sol}
\ee
As the RHS of the above equation decreases from 1 when $\va_{\nu_0}/\Omega\simeq 0_-$, to $\Omega/(-3\va_{\nu_0})$ when $\va_{\nu_0}/\Omega \rightarrow  -\infty$, such a separable potential has a single bound state solution provided that $v_0$ is larger than a threshold value $v_{th}=1/2\rho_0$ obtained by setting $\va_{\nu_0}=0$ in the above equation. Let us write the bound state solution of Eq.~(\ref{eq:bounds_sol}) as
\begin{subeqnarray}
\tilde \va_{\nu_0}&\equiv&\frac{\va_{\nu_0}}{\Omega}=f(\tilde v_0)\, ,\\
\tilde v_0&\equiv&\frac{v_0}{v_{th}}=2\rho_0v_0\, \slabel{eq:tildev0=vo}.
\end{subeqnarray}

Equation (\ref{eq:vpinui}) gives the normalized bound state wave function as
\be
\lan \vp|\nu_0\ran=\frac{w_\vp}{\va_\vp-\va_{\nu_0}}\frac{1}{\sqrt{I_2}}\label{eq:wavefgs}
\ee
where
\be
I_n=\sum_\vp \frac{w_\vp}{(\va_\vp-\va_{\nu_0})^n}
\ee
obeys a recursion relation derived from an integration by parts. It reads
\be
0=(2n-3)I_n +2n\va_{\nu_0} I_{n+1}+\frac{2\rho_0}{\Omega^{n-1}(1-\tilde \va_{\nu_0})^n}\, .\label{eq:recursionre}
\ee
Knowing $I_1=1/v_0$ and $2\rho_0=1/v_{th}$ according to Eqs.~(\ref{eq:singlepaireq}) and (\ref{eq:tildev0=vo}), this recursion relation gives
\be
I_2=\frac{1}{-2\va_{\nu_0}v_0}\left(\frac{\tilde v_0 }{1-\tilde \va_{\nu_0} }-1\right)\, .
\ee

{\bf $\bullet$ Extended states} \

\mbox{}

Equation (\ref{eq:schroeq_1dimer}) also has nonnegative solutions. The corresponding $\va_{\nu_i}$'s are close to the possible $\va_{\vp_i}$ energies of free pair states. The resulting wave function $\lan \vp|\nu_i\ran$ then is very much peaked at $\va_\vp=\va_{\vp_i}$, which makes the extended states very close to the free plane-wave states for all $\vp_i$'s within the potential layer.
%

\subsection{Elementary scatterings between two ground-state cobosons}

Knowing the single-pair ground-state wave function, we can calculate a few important quantities of the coboson many-body formalism dealing with ground states. Let us start with the simplest ones, namely, the Pauli scattering $\lambda\left(\begin{smallmatrix} 0& 0\\ 0& 0 \end{smallmatrix}\right)$ and the interaction scattering $\xi\left(\begin{smallmatrix} 0& 0\\ 0& 0 \end{smallmatrix}\right)$ between two cobosons starting and ending in the ground state, $0=(\vK=0,\nu_0)$.\

\begin{figure}[h!]
\begin{center}
\includegraphics[trim=6cm 6.8cm 6cm 19.2cm,clip,width=2.7in] {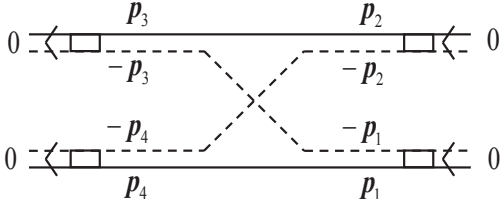}
\end{center}
\vspace{-0.3cm}
\caption{{\small Pauli scattering $\lambda\big(^{\hspace{.05cm}
0 \hspace{.09cm} 0\hspace{.05cm}}_{\hspace{.05cm}
0\hspace{.09cm} 0\hspace{.05cm}}\big)$ between ground-state cobosons $0=(\vK=0,\nu_0)$, as given in Eq.~(\ref{eq:lambda0000}). Momentum conservation for fermion exchange imposes all $\vp$'s to be equal. }}
\label{fig:Pauliscattxi0000}
\end{figure}

(i) The Pauli scattering $\lambda\left(\begin{smallmatrix} n& j\\ m& i \end{smallmatrix}\right)$ is defined in Eq.~(\ref{def_lambda}) and visualized by the Shiva diagram of Fig.~\ref{fig:lambda_mnij}. According to Eq.~(\ref{eq:defBKp}), the momenta of fermions $(\alpha,\beta)$ making a coboson with center-of-mass momentum $\vK=0$ reduce to $(\vp,-\vp)$; so, the overlap $\lan \vk_\alpha ,\vk_\beta|\nu_0\ran$ between the free fermion pair and the $(\vK=0,\nu_0)$ coboson is just $\lan \vp|\nu_0\ran$. The Pauli scattering $\lambda\left(\begin{smallmatrix} 0& 0\\ 0& 0 \end{smallmatrix}\right)$ between two ground-state cobosons then is simply given, using Eq.~(\ref{eq:wavefgs}), by
\be
\lambda\left(\begin{smallmatrix} 0& 0\\ 0& 0 \end{smallmatrix}\right)=\sum_\vp |\lan \vp|\nu_0\ran|^4=\frac{I_4}{I_2^2}\, .\label{eq:lambda0000}
\ee
Indeed, as directly read on Fig.~\ref{fig:Pauliscattxi0000}, the diagram that represents $\lambda\left(\begin{smallmatrix} 0& 0\\ 0& 0 \end{smallmatrix}\right)$ imposes all momenta $\vp$'s to be equal, as can also be shown from Eq.~(\ref{def_lambda}).
Using $I_4$ calculated through the recursion relation (\ref{eq:recursionre}) as
\be
I_4=\frac{1}{-48\va^3_{\nu_0}v_0}\left(\tilde v_0\frac{3-12\tilde \va_{\nu_0} +17\tilde \va^2_{\nu_0}}{(1-\tilde \va_{\nu_0})^3}-3\right),
\ee
we end with
\be
\lambda\left(\begin{smallmatrix} 0& 0\\ 0& 0 \end{smallmatrix}\right)=\frac{v_0}{-12  \va_{\nu_0}}\frac{\tilde v_0\displaystyle \frac{3-12\tilde \va_{\nu_0} +17\tilde \va^2_{\nu_0}}{(1-\tilde \va_{\nu_0})^3}-3}{\left(\displaystyle \frac{\tilde v_0}{1- \tilde \va_{\nu_0} }-1\right)^2}\, .
\ee
We note that $\lambda\left(\begin{smallmatrix} 0& 0\\ 0& 0 \end{smallmatrix}\right)$ is dimensionless as required, and depends on the sample volume as $1/L^3$ due to the volume dependence of the $v_0$ amplitude. \

\begin{figure}[h]
\begin{center}
\includegraphics[trim=4cm 6.4cm 4.7cm 19.2cm,clip,width=3.2in] {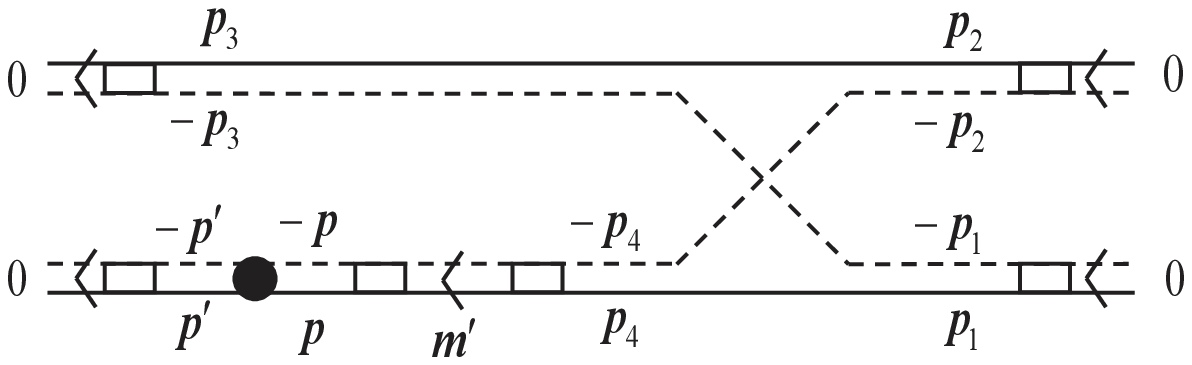}
\end{center}
\vspace{-0.4cm}
\caption{{\small Interaction scattering $\xi\big(^{\hspace{.05cm}
0 \hspace{.09cm} 0\hspace{.05cm}}_{\hspace{.05cm}
0\hspace{.09cm} 0\hspace{.05cm}}\big)$ between ground-state cobosons, as given in Eqs.~(\ref{eq:xi0000}) and (\ref{eq:xi0000_2}). }}
\label{fig:interactionscattxi0000}
\end{figure}

(ii) The interaction scattering $\xi\left(\begin{smallmatrix} n& j\\ m& i \end{smallmatrix}\right)$ is defined in Eq.~(\ref{eq:xi_mnij}) and visualized by the diagram of Fig.~\ref{fig:Xi_mnij}. For $i=j=m=n=0$, all four terms in $\xi\left(\begin{smallmatrix} n& j\\ m& i \end{smallmatrix}\right)$ are equal, thus giving
\be
\xi\left(\begin{smallmatrix} 0& 0\\ 0& 0 \end{smallmatrix}\right)=2\sum_{m'}v_{0m'}\lambda\left(\begin{smallmatrix} 0& 0\\ m'& 0 \end{smallmatrix}\right)\, .\label{eq:xi0000}
\ee
Since $v_{0m'}$ conserves momentum, all cobosons involved in this scattering have a $\vK=0$ center-of-mass momentum; so, they are all made of $(\vp,-\vp)$ fermion pairs. The above sum is visualized by the Shiva diagram of Fig.~\ref{fig:interactionscattxi0000}.
The Pauli scattering $\lambda\left(\begin{smallmatrix} 0& 0\\ m'& 0 \end{smallmatrix}\right)$ imposes $\vp_1=\vp_2=\vp_3=\vp_4$, while the sum over $m'$, performed through closure relation, $\sum |m'\ran\lan m'|={\rm I}$, leads to $\vp=\vp_4$. So, the interaction scattering $\xi\left(\begin{smallmatrix} 0& 0\\ 0& 0 \end{smallmatrix}\right)$, as directly read from this diagram, is given by
\bea
\xi\left(\begin{smallmatrix} 0& 0\\ 0& 0 \end{smallmatrix}\right)&=&2\sum_{\vp'\vp} \lan \nu_0|\vp'\ran v_{\vp'-\vp}\lan \nu_0|\vp\ran\lan \vp|\nu_0\ran^2\nn\\
&=&2v_0\frac{I_1I_3}{I^2_2}=2\frac{I_3}{I^2_2}\, ,\label{eq:xi0000_2}
\eea
since $I_1=1/v_0$. It is of interest to note that $2I_3/I_2^2$ holds true even for a non-separable potential, as can be shown by inserting the single coboson Schr\"{o}dinger equation (\ref{eq:schroeq_1dimer}) for $\nu_i=\nu_0$ into the above equation. Using $I_3$ calculated through the recursion relation (\ref{eq:recursionre}) as
\be
I_3=\frac{1}{8  \va^2_{\nu_0} v_0}\left(\tilde v_0 \frac{1-3 \tilde \va_{\nu_0}}{(1-\tilde \va_{\nu_0})^2}-1\right)\, ,
\ee
we end with
\be
\xi\left(\begin{smallmatrix} 0& 0\\ 0& 0 \end{smallmatrix}\right)=v_0\frac{\displaystyle \tilde v_0 \frac{1-3 \tilde \va_{\nu_0}}{(1-\tilde \va_{\nu_0})^2}-1}{\left(  \displaystyle\frac{\tilde v_0}{1-\tilde \va_{\nu_0}}-1\right)^2}\, .
\ee

\begin{figure}[h]
\begin{center}
\includegraphics[trim=5cm 7.2cm 5cm 19cm,clip,width=3in] {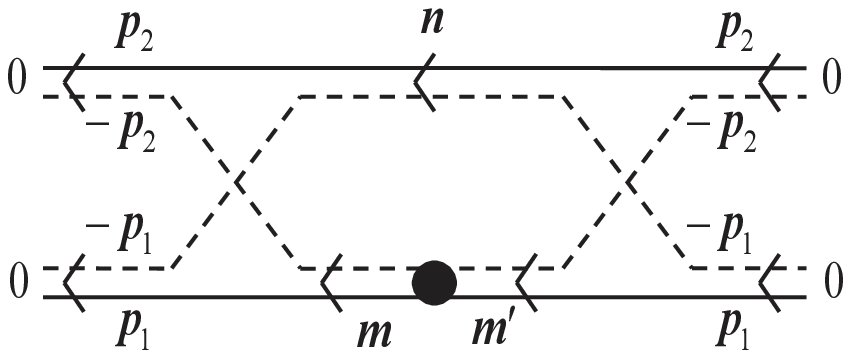}
\end{center}
\vspace{-0.3cm}
\caption{{\small Interaction scattering $\xi^{in}\big(^{\hspace{.05cm}
0 \hspace{.09cm} 0\hspace{.05cm}}_{\hspace{.05cm}
0\hspace{.09cm} 0\hspace{.05cm}}\big)$ between ground-state cobosons, as given in Eqs.~(\ref{eq:xiinscattering},\ref{eq:chiinscattering}). }}
\label{fig:interactionscattxiin0000}
\end{figure}

(iii) The interaction scattering $\xi^{in}\left(\begin{smallmatrix} n& j\\ m& i \end{smallmatrix}\right)$ is defined in Eq.~(\ref{eq:xiin_mnij}) and visualized by the diagram of Fig.~\ref{fig:Xiin_mnij}. Here again, since all scatterings conserve momentum, all cobosons involved in $\xi^{in}\left(\begin{smallmatrix} 0& 0\\ 0& 0 \end{smallmatrix}\right)$ have a $\vK=0$ center-of-mass momentum. Moreover, the various contributions to $\lambda\left(\begin{smallmatrix} 0& n\\ 0& m \end{smallmatrix}\right)$ and $\xi\left(\begin{smallmatrix} n& 0\\ m& 0 \end{smallmatrix}\right)$ are equal; so, $\xi^{in}\left(\begin{smallmatrix} 0& 0\\ 0& 0 \end{smallmatrix}\right)$ reduces to
\be
\xi^{in}\left(\begin{smallmatrix} 0& 0\\ 0& 0 \end{smallmatrix}\right)=2\sum_{mm'n} \lambda\left(\begin{smallmatrix} 0& n\\ 0& m \end{smallmatrix}\right) v_{mm'} \lambda\left(\begin{smallmatrix} n& 0\\ m'& 0 \end{smallmatrix}\right)\, .\label{eq:xiinscattering}
\ee
The above sum is visualized by the Shiva diagram of Fig.~\ref{fig:interactionscattxiin0000}.
By performing the $(m,m',n)$ sums through closure relation, we are left with two free momenta only, $\xi^{in}\left(\begin{smallmatrix} 0& 0\\ 0& 0 \end{smallmatrix}\right)$ then reading as
\bea
\xi^{in}\left(\begin{smallmatrix} 0& 0\\ 0& 0 \end{smallmatrix}\right)&=&2\sum_{\vp_1} |\lan \vp_1|\nu_0\ran|^2 \sum_{\vp_2}v_{\bf0} |\lan \vp_2|\nu_0\ran|^2 \nn\\
&=&2v_0\,.\label{eq:chiinscattering}
\eea

\subsection{Scattering length}
In this section, we make link between the above coboson-coboson scatterings and the coboson-coboson scattering length. We focus on systems that have physical relevance to cold atom gases, namely 3D systems with a potential energy layer extending from $0$ to $\Omega$. As previously shown, a bound state exists for a separable potential provided that $v_0 $ is larger than a threshold value $ v_{th}$. Let us consider two limiting cases: (i) $\va_{\nu_0}/\Omega$ close to zero, and (ii) $\va_{\nu_0}/\Omega$ infinite. \

 (i) For $\va_{\nu_0}/\Omega \rightarrow 0_-$, Eq.~(\ref{eq:bounds_sol}) gives,
\be
\frac{\pi}{2}\sqrt{\frac{-\va_{\nu_0}}{\Omega}}\simeq\frac{\tilde v_0-1}{\tilde v_0} \, ,\label{eq:vanu0v0case1}
\ee
since $\tan^{-1} x=\pi/2$ for $x\rightarrow \infty$. The solution for $\va_{\nu_0}/\Omega \rightarrow 0_-$ corresponds to $\tilde v_0\sim 1$, i.e., $v_0 \simeq v_{th}=1/2\rho_0$. The coboson ground state is then loosely bound. Such a regime is physically relevant for cold atom gases. \

By noting that the density of states is defined through $(L/2\pi)^34\pi k^2dk=\rho_0\sqrt{\va_\vk/\Omega}d\va_\vk$ for $\va_\vk=k^2/2\mu$ with $\mu^{-1}=m_\alpha^{-1}+m_\beta^{-1}$, we obtain
\be
\frac{L^3}{2\pi^2}=\frac{2\rho_0}{\sqrt{\Omega}(2\mu)^{3/2}}\, .\label{eq:relLrhoomega}
\ee
Using this relation in Eq.~(\ref{eq:vanu0v0case1}), we get, for a bound-state energy written in terms of the bound-state Bohr radius $a$ as $\va_{\nu_0}=-1/2\mu a^2$,
\be
\frac{v_0}{\tilde v_0-1}=\frac{2\pi}{\mu L^3}a\, .\label{eq:tildev_0a}
\ee
The above relation gives the Pauli and interaction scatterings for $\va_{\nu_0}/\Omega \rightarrow 0_-$, i.e., $\tilde v_0\sim 1$, as
\bea
\lambda\left(\begin{smallmatrix} 0& 0\\ 0& 0 \end{smallmatrix}\right) &\simeq&\frac{1}{-4\va_{\nu_0}}\frac{v_0}{\tilde v_0-1}=\pi \left(\frac{a}{L}\right)^3\, ,\\
\xi\left(\begin{smallmatrix} 0& 0\\ 0& 0 \end{smallmatrix}\right)&\simeq& \frac{v_0}{\tilde v_0-1}= \frac{2\pi}{\mu L^3}a \gg \xi^{in}\left(\begin{smallmatrix} 0& 0\\ 0& 0 \end{smallmatrix}\right)=2v_0\, .\label{eq:xia}
\eea
Thus, in this regime, the interaction scattering is dominantly controlled by fermion exchange.\

As shown below, the coboson-coboson effective scattering appears in the Born approximation as $\xi\left(\begin{smallmatrix} 0& 0\\ 0& 0 \end{smallmatrix}\right)-\xi^{in}\left(\begin{smallmatrix} 0& 0\\ 0& 0 \end{smallmatrix}\right)$. By relating this effective scattering to the coboson-coboson scattering length $a^{(B)}_{cc}$ through\cite{Fetterbook}
\be
\xi\left(\begin{smallmatrix} 0& 0\\ 0& 0 \end{smallmatrix}\right)-\xi^{in}\left(\begin{smallmatrix} 0& 0\\ 0& 0 \end{smallmatrix}\right)=\frac{4\pi}{ML^3}a^{(B)}_{cc}\, ,
\ee
we get, from Eqs.~(\ref{eq:chiinscattering}) and (\ref{eq:xia}), this scattering length as
\be
a^{(B)}_{cc}\simeq\frac{M}{2\mu}a\, ,
\ee
which reduces to $2a$ for equal fermion masses, $m_\alpha=m_\beta$. This Born result has been previously found by other many-body approaches\cite{Haussmann1993,Melo1993,PieriPRB2000}. It has also been shown that the repeated interaction through coboson-coboson ladder-type series drastically reduces this Born result down to $0.6a$ for equal fermion masses\cite{Brodsky2005,levinsen2011,Alzetto2013,Petrov2004}. Derivation of this reduction using the present coboson formalism is left for future work.  \

(ii) The $\va_{\nu_0}/\Omega \rightarrow -\infty$ regime corresponds to an attractive potential with infinite strength, $\tilde v_0\rightarrow \infty$, and a tightly bound coboson with a very large binding energy. Equation (\ref{eq:bounds_sol}) then gives in this regime
\be
\va_{\nu_0}=-v_0\frac{2\rho_0\Omega}{3}=-v_0 N_\Omega\, ,
\ee
where $N_\Omega=\sum_\vk w_\vk$ is the number of states in the potential layer. The Pauli and interaction scatterings are given in this limit by
\bea
\lambda\left(\begin{smallmatrix} 0& 0\\ 0& 0 \end{smallmatrix}\right) &\simeq&\frac{3}{2}\frac{1}{\rho_0\Omega}=\frac{1}{N_\Omega}\, ,\\
\xi\left(\begin{smallmatrix} 0& 0\\ 0& 0 \end{smallmatrix}\right)&\simeq& 2v_0\, .\label{eq:xiv0}
\eea
These results show that the Pauli scattering $\lambda\left(\begin{smallmatrix} 0& 0\\ 0& 0 \end{smallmatrix}\right)$ between tightly bound cobosons is vanishingly small as physically reasonable, while the coboson-coboson effective scattering in the Born approximation, $\xi\left(\begin{smallmatrix} 0& 0\\ 0& 0 \end{smallmatrix}\right)-\xi^{in}\left(\begin{smallmatrix} 0& 0\\ 0& 0 \end{smallmatrix}\right)$, reduces to zero.

\subsection{Normalization factor}

Let us now apply the coboson formalism developed above for interaction between fermions $(\alpha, \beta)$ only to the calculation of two specific quantities: the normalization factor for $N$ ground-state cobosons and the Hamiltonian mean value in this state. The former quantity measures to what extent cobosons entangle with each other; the latter corresponds to the $N$-coboson ground-state energy in the Born approximation. These two quantities have already been calculated for semiconductor excitons with Coulomb potential acting between all carriers\cite{M_EPJC2003,Odile_EPJB2003}. Explicit results will be given in the weak interaction regime, which is the regime relevant for cold atom gases.

Due to the Pauli exclusion principle, the scalar product of the $N$-coboson state $|\psi_N\ran= B^{\dag N}_0|v\ran$ with itself is far smaller than its elementary boson value $N!$. Let us write it as
\be
\lan \psi_N|\psi_N\ran=N!F_N\, .
\ee
Following the general coboson many-body formalism, we get, by iterating Eqs.~(\ref{eq:commut_BB}) and (\ref{eq:commut_DB}),
\bea
\left[B_m, B^{\dag N}_i\right]_-&=& N B^{\dag N-1}_i (\delta_{mi}-D_{mi}) \nn\\
&&-N(N-1) B^{\dag N-2}_i\sum_nB^\dag_n \lambda\big(\begin{smallmatrix} n& i\\ m& i \end{smallmatrix}\big)\, ,\label{eq:BBN_commu}\\
\left[D_{mi}, B^{\dag N}_j\right]_-&=& N B^{\dag N-1}_i  \sum_nB^\dag_n \Big(\lambda\big(\begin{smallmatrix} n& j\\ m& i \end{smallmatrix}\big)+(i\longleftrightarrow j)\Big)\, .\nn\\
\label{eq:DBN_commu}
\eea
The recursion relation between the $F_N$'s is obtained by using the above equations in
\bea
\lan \psi_N|\psi_N\ran&=& \lan \psi_{N-1}|\Big[B_0,B^{\dag N}_0\Big]_-|v\ran\\
&=& N\lan \psi_{N-1}|\psi_{N-1}\ran\nn\\
&&-N(N-1)\sum_n \lambda\big(\begin{smallmatrix} n& 0\\ 0& 0 \end{smallmatrix}\big)\lan \psi_{N-1}|B^\dag_n|\psi_{N-2}\ran\, ,\nn
\eea
This leads to
\be
F_N=F_{N-1}-(N-1)\lambda_2 F_{N-2}+(N-1)(N-2)\lambda_3F_{N-3}+\cdots\, .\label{eq:recursionFN}
\ee
The $\lambda_n$'s, which come from fermion exchanges between $n$ cobosons $0$, reduce, for the ground-state coboson wave function $\lan \vp|\nu_0\ran$ given in Eq.~(\ref{eq:wavefgs}), to
\be
\lambda_n=\sum_\vp |\lan \vp|\nu_0\ran|^{2n}= \frac{I_{2n}}{I_2^n}\, .
\ee
$\lambda_2$ is just $\lambda\big(\begin{smallmatrix} 0& 0\\ 0& 0 \end{smallmatrix}\big)$ given in Eq.~(\ref{eq:lambda0000}), while $\lambda_3=\lambda\left(\begin{smallmatrix} 0& 0\\ 0& 0 \\ 0& 0  \end{smallmatrix}\right)$ and so on. The $\lambda_n$'s can be derived using the recursion relation (\ref{eq:recursionre}). For $\va_{\nu_0}/\Omega \rightarrow 0_-$, that is, in the weak interaction limit $\tilde v_0\rightarrow 1$, we find that

\be
I_n=\frac{(2n-5)!!}{(n-1)!(-2\va_{\nu_0})^{n-1}}\frac{\tilde v_0-1}{v_0}\, \mbox{ for } n>2 \, .
\ee
The above result is obtained by using Eq.~(\ref{eq:tildev_0a}) and by noting that the third term in Eq.~(\ref{eq:recursionre}) is small compared to the other two terms; so, $I_n$ becomes a geometric series. With the use of the above relation, the $\lambda_n$'s read in terms of the coboson Bohr radius $a$ as
\be
\lambda_n=\frac{(2\pi)^{n-1}(4n-5)!!}{(2n-1)!}\left(\frac{a^3}{L^3}\right)^{n-1}\, .
\ee

\subsection{Hamiltonian mean value}

The Hamiltonian mean value for $N$ ground-state cobosons $|\psi_N\ran$ corresponds to the ground-state energy in the Born approximation, since, by construction, it contains one interaction only. However, for composite quantum particles, the Hamiltonian mean value also contains fermion exchanges between $N$ cobosons; so, it also has terms with higher than linear density dependence.\

\begin{figure}[h]
\begin{center}
\includegraphics[trim=2.8cm 2.4cm 2cm 15.3cm,clip,width=3.5in] {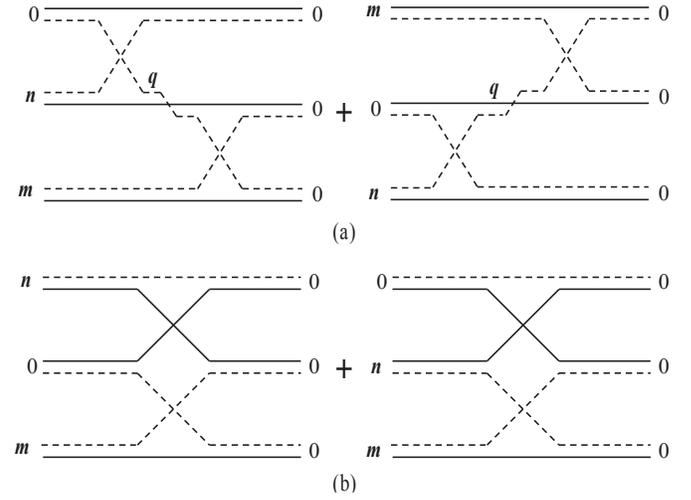}
\end{center}
\vspace{-0.3cm}
\caption{{\small (a) Shiva diagrams for the two terms of the $q$ sum in Eq.~(\ref{eq:lmnp}). (b) The two diagrams of (a) do correspond to fermion exchanges involving cobosons $(m,n)$ in a symmetrical way.}}
\label{fig:Shivadiaeq64}
\end{figure}

  Iteration of Eqs.~(\ref{eq:HB_commu}) and (\ref{eq:ViBjxi}) leads to
 \bea
 \Big[H,B^{\dag N}_i\Big]_-&=& NB_i^{\dag N-1}(E_i B^\dag_i+V^\dag_i) \\
 &&+\frac{N(N-1)}{2}B_i^{\dag N-2}\sum_{mn} B^\dag_mB^\dag_n \xi\left(\begin{smallmatrix} n& i\\ m& i \end{smallmatrix}\right)\, ,\nn\\
 \Big[V^\dag_i,B_j^{\dag N}\Big]_-&=& NB_j^{\dag N-1} \sum_{mn} B^\dag_mB^\dag_n \xi\left(\begin{smallmatrix} n& j\\ m& i \end{smallmatrix}\right)\, ,
 \eea
from which we find
\be
H|\psi_N\ran= NE_0|\psi_N\ran +\frac{N(N-1)}{2}\sum_{mn} B^\dag_mB^\dag_n \xi\left(\begin{smallmatrix} n& 0\\ m& 0 \end{smallmatrix}\right)|\psi_{N-2}\ran\, ,
\ee
where $E_0$ is just the coboson ground-state energy $\va_{\nu_0}$. So, we are left with calculating
\bea
\lefteqn{\lan \psi_N|H-NE_0|\psi_N\ran=}\label{eq:H-NE0}\\
&&\,\,\,\,\,\,\,\frac{N(N-1)}{2}\sum_{mn}  \xi\left(\begin{smallmatrix} n& 0\\ m& 0 \end{smallmatrix}\right)\lan \psi_N|B^\dag_mB^\dag_n|\psi_{N-2}\ran\, \nn.
\eea
This is done by using Eqs.~(\ref{eq:BBN_commu}) and (\ref{eq:DBN_commu}). These equations give
\bea
\lefteqn{B_mB_n|\psi_N\ran=N(N-1)l_{mn}|\psi_{N-2}\ran}\label{eq:BmnpsiN}\\
&&-N(N-1)(N-2)\sum_p l_{mnp} B^\dag_p|\psi_{N-3}\ran\nn\\
&&+N(N-1)(N-2)(N-3)\sum_{pq}l_{mnpq} B^\dag_p B^\dag_q  |\psi_{N-4}\ran\, ,\nn
\eea
where
\bea
l_{mn}&=& \delta_{m0}\delta_{n0}-\lambda\left(\begin{smallmatrix} n& 0\\ m& 0 \end{smallmatrix}\right)\, , \\
l_{mnp}&=&\delta_{m0} \lambda\left(\begin{smallmatrix} p& 0\\ n& 0 \end{smallmatrix}\right)+\delta_{n0} \lambda\left(\begin{smallmatrix} p& 0\\ m& 0 \end{smallmatrix}\right)\nn\\
&&-\sum_q  \Big(\lambda\left(\begin{smallmatrix} p& 0\\ n& q \end{smallmatrix}\right)+\lambda\left(\begin{smallmatrix} p& q\\ n& 0 \end{smallmatrix}\right)\Big)\lambda\left(\begin{smallmatrix} q& 0\\ m& 0 \end{smallmatrix}\right)\, ,\label{eq:lmnp}\\
l_{mnpq}&=& \lambda\left(\begin{smallmatrix} p& 0\\ m& 0 \end{smallmatrix}\right)\lambda\left(\begin{smallmatrix} q& 0\\ n& 0 \end{smallmatrix}\right)\, .
\eea
Terms higher than $N(N-1)(N-2)(N-3)$ can be obtained by iterating Eq.~(\ref{eq:BmnpsiN}). Using this equation, we get
\be
\lan \psi_{N-2}|B_mB_n|\psi_N\ran\simeq N!\Big(F_{N-2}l_{mn}-(N-2)F_{N-3}l_{mn0}\Big)\label{eq:N-2BmBnN}
\ee
within terms in $(N-2)(N-3)F_{N-4}$. The $q$ sum in $l_{mn0}$ does not look symmetrical at first with respect to $(m,n)$. Yet, the Shiva diagrams in Fig.~\ref{fig:Shivadiaeq64}(a) show that the two terms of this sum correspond to two exchange processes in which cobosons $(m,n)$ play an equivalent role, a fact which becomes obvious from Fig.~\ref{fig:Shivadiaeq64}(b).

Equation (\ref{eq:N-2BmBnN}) used in Eq.~(\ref{eq:H-NE0}) gives
\bea
\lefteqn{\frac{\lan \psi_N|H-N\va_{\nu_0}|\psi_N\ran}{N!}=}\nn\\
&&\frac{N(N-1)}{2}F_{N-2}\sum_{mn}\Big(\delta_{0m}\delta_{0n}-\lambda\left(\begin{smallmatrix} 0& n\\ 0& m \end{smallmatrix}\right)\Big)\xi\left(\begin{smallmatrix} n& 0\\ m& 0 \end{smallmatrix}\right)\nn\\
&&-\frac{N(N-1)}{2}(N-2)F_{N-3}\sum_{mn}\bigg(\delta_{0m}\lambda\left(\begin{smallmatrix} 0& 0\\ 0& n \end{smallmatrix}\right)+\delta_{0n}\lambda\left(\begin{smallmatrix} 0& 0\\ 0& m \end{smallmatrix}\right)\nn\\
&&-\lambda\left(\begin{smallmatrix} 0& 0\\ 0& n  \\ 0& m \end{smallmatrix}\right)-\lambda\left(\begin{smallmatrix} 0& n\\ 0& 0 \\ 0& m \end{smallmatrix}\right)\bigg)\xi\left(\begin{smallmatrix} n& 0\\ m& 0 \end{smallmatrix}\right)+\cdots\, .\label{eq:H-NE0N!}
\eea

\begin{figure}[t!]
\begin{center}
\includegraphics[trim=2.5cm 5.4cm 2cm 18cm,clip,width=3.5in] {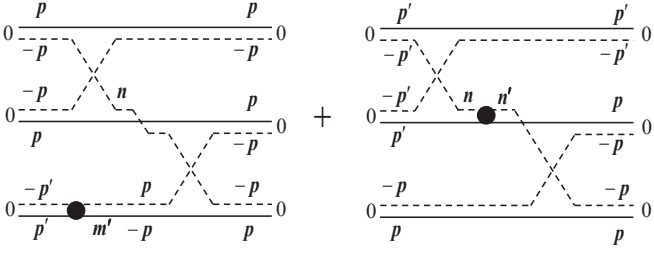}
\end{center}
\vspace{-0.4cm}
\caption{{\small Diagrams corresponding to the two terms in Eq.~(\ref{eq:Shivadiaeq642}).}}
\label{fig:Shivadiaeq642}
\end{figure}

\begin{figure}[h!]
\begin{center}
\includegraphics[trim=2.1cm 2cm 1cm 15cm,clip,width=3.6in] {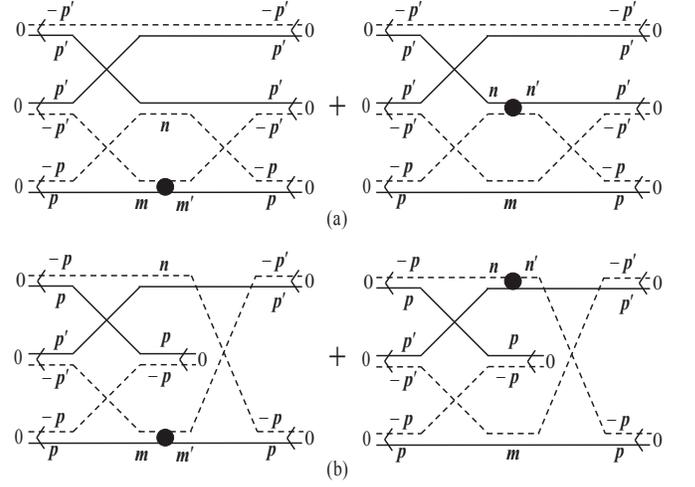}
\end{center}
\vspace{-0.4cm}
\caption{{\small Interaction scatterings involving fermion exchange between three cobosons $0$. The scatterings in (a) or (b) lead to Eq.~(\ref{eq:Shivadiaeq6432}).}}
\label{fig:Shivadiaeq6432}
\end{figure}

The first $(m,n)$ sum reduces to $\xi\left(\begin{smallmatrix} 0& 0\\ 0& 0 \end{smallmatrix}\right)-\xi^{in}\left(\begin{smallmatrix} 0& 0\\ 0& 0 \end{smallmatrix}\right)$, which is the coboson-coboson effective scattering in the Born approximation, as previously mentioned. The first two terms of the second $(m,n)$ sum, visualized by the Shiva diagrams of Fig.~\ref{fig:Shivadiaeq642}, are equal. By summing over $(n,m')$ and $(n,n')$ through closure relations, we get
\bea
\lefteqn{\sum_{mn} \delta_{0m}\lambda\left(\begin{smallmatrix} 0& 0\\ 0& n \end{smallmatrix}\right)\xi\left(\begin{smallmatrix} n& 0\\ m& 0 \end{smallmatrix}\right)=\sum_{\vp\vp'} \lan \nu_0|\vp'\ran  \lan \nu_0|\vp\ran^2 v_{\vp'-\vp}\lan \vp|\nu_0\ran^3}\nn\\
&&+ \sum_{\vp\vp'} \lan \nu_0|\vp'\ran^2  \lan \nu_0|\vp\ran  v_{\vp'-\vp} \lan \vp'|\nu_0\ran\lan \vp|\nu_0\ran^2\hspace{1.7cm}\nn\\
&&= v_0\frac{I_1I_5}{I_2^3}+v_0\frac{I_3^2}{I_2^3}\,.\label{eq:Shivadiaeq642}
\eea
The other two terms of the second sum in Eq.~(\ref{eq:H-NE0N!}) involve fermion exchange between three cobosons. They are shown in Fig.~\ref{fig:Shivadiaeq6432}. The term in $\lambda\left(\begin{smallmatrix} 0& 0\\ 0& n  \\ 0& m \end{smallmatrix}\right)$, shown in Fig.~\ref{fig:Shivadiaeq6432}(a), leads to
\be
2\sum_{\vp\vp'} \lan \nu_0|\vp\ran \lan \nu_0|\vp'\ran^2  v_{\bf0} \lan \vp| \nu_0\ran\lan \vp'|\nu_0\ran^2=2v_0\frac{I_4}{I_2^2}\, ,\label{eq:Shivadiaeq6432}
\ee
with an equal contribution from the term in $\lambda\left(\begin{smallmatrix} 0& n\\ 0& 0 \\ 0& m \end{smallmatrix}\right)$, shown in Fig.~\ref{fig:Shivadiaeq6432}(b). Note that in these processes, the interaction is associated with zero momentum transfer, but the $(m,m')$ and $(n,n')$ cobosons can have a center-of-mass momentum that differs from $0$.\

Using the above results, we find the Hamiltonian mean value as
\bea
\lan H\ran_N&=&\frac{\lan \psi_N|H|\psi_N\ran}{\lan \psi_N|\psi_N\ran}\\
&\simeq&N\va_{\nu_0}+\frac{N(N-1)}{2}2v_0\bigg\{\frac{F_{N-2}}{F_N}\left( \frac{I_1I_3}{I_2^2}-1\right)\nn\\
&&-(N-2)\frac{F_{N-3}}{F_N}\left(\frac{I_1I_5+I_3^2}{I_2^3}-\frac{2I_4}{I_2^2}\right)\bigg\}\, .\nn
\eea
In the large $N$ limit, the recursion relation (\ref{eq:recursionFN}) between the $F_N$'s gives the $F_N$ ratios as
\bea
\frac{F_{N-2}}{F_N}&=&\frac{F_{N-2}}{F_{N-1}}\frac{F_{N-1}}{F_N}\simeq\left(\frac{F_{N-1}}{F_N}\right)^2\nn\\
&\simeq& 1+2N\lambda_2\,
\eea
up to first order in density. So, using Eq.~(\ref{eq:lambda0000}) for $\lambda_2$, we end with
\bea
\frac{\lan H\ran_N}{N}&\simeq&\va_{\nu_0}+ Nv_0 \left( \frac{I_1I_3}{I_2^2}-1\right)\nn\\
&&+ N^2 v_0 \left(2\frac{I_1I_3I_4}{I_2^4}-\frac{I_1I_5+I_3^2}{I_2^3}\right)+\cdots\, .
\eea
In the weak interaction limit $\tilde v_0\rightarrow 1$, which is the regime relevant for cold atom gases, we find that the Hamiltonian mean value can be written in terms of the dimensionless many-body parameter $\eta=N(a/L)^3$ as
\be
\lan H\ran_N\simeq N|\va_{\nu_0}|\left(-1+2\pi \eta-\pi^2\eta^2+\cdots\right)\, ,\label{eq:DensityHMV}
\ee
within small corrections in $(v_0L^3/a^3)\propto \tilde v_0-1\simeq 0$, as seen from Eq.~(\ref{eq:xia}).\

In the case of Cooper pairs, the reduced BCS potential
\be
V_{BCS}=-v_0\sum_{\vk\vk'}w_{\vk}w_{\vk'} a^\dag_{\vk'\uparrow}a^\dag_{-\vk'\downarrow}a_{-\vk\downarrow}a_{\vk\uparrow}
\ee
also is a one-body operator in the fermion pair subspace\cite{M_EPJB2011}; so, many-body effects between Cooper pairs also are entirely controlled by fermion exchange. However, Pauli blocking induced by the Pauli exclusion principle prohibits interaction between more than two Cooper pairs because the pairs interacting through $V_{BCS}$ have a zero center-of-mass momentum only. This makes the density dependence of the $N$-Cooper pair ground-state energy reduce to a linear term\cite{M_EPJB2011_2,Crouzeix2011}. The nonlinear density dependence of $\lan H\ran_N$ obtained in Eq.~(\ref{eq:DensityHMV}) comes from the fact that we here deal with cobosons having two quantum indices: a center-of-mass momentum and a relative-motion momentum. They come from the $(\vk_\alpha,\vk_\beta)$ fermion pairs from which these cobosons are made, in contrast to Cooper pairs, which are single-index objects made of $(\vp,-\vp)$ pairs.


\section{Conclusion}

This paper presents the coboson many-body formalism appropriate for a specific set of problems, namely two-fermion coboson gas in which the interaction is restricted to attraction between different fermion species. Such a potential, commonly used in two major fields of physics, namely BCS superconductors and cold atom gases, reads as a one-body operator in the coboson subspace. This feature has an important consequence: coboson many-body effects then are entirely driven by fermion exchanges, making their Shiva diagrams considerably simpler than those for semiconductor excitons with Coulomb interaction between all carriers. Analytical results are given for potential scattering having a separable form, as commonly taken when the physics at hand is controlled by short-range interaction. We have here studied two limiting cases in detail, the strong and the weak interaction regimes. In the strong interaction regime in which the cobosons are tightly bound fermion pairs, the effective coboson-coboson scattering cancels, while the Pauli scattering brings vanishingly small effect. These cobosons appear as (nearly) noninteracting bosons. By contrast, in the weak interaction regime, the dominant many-body effects come from interactions through fermion exchange, as visualized by Shiva diagrams. We here recover the well-known Born result, $2a$, for the dimer-dimer scattering length for cold atoms having same mass. Calculation of the dimer-dimer scattering length beyond Born approximation using the coboson many-body formalism presented here will be explored in a future work. We end this work by calculating the normalization factor and the Hamiltonian mean value in the $N$-ground-state-coboson state, the latter corresponding to the ground-state energy in the Born approximation. The Hamiltonian mean value has terms nonlinear in density that come from fermion exchange between more than two cobosons, although only one potential scattering occurs in the Hamiltonian mean value by construction. The appearance of such nonlinear terms is standard for systems made of cobosons having two degrees of freedom such as semiconductor excitons.\

 The advantage of the coboson many-body formalism over standard elementary fermion approaches is two-fold: (i) At the single pair level, a complete set of coboson states, including bound and extended states, is introduced; thus, at the very beginning of the formalism, we are exempt from summing infinite ladder series in order to reach the singular poles corresponding to coboson bound states. (ii) At the many-body level, the state in which all $N$ cobosons are in the ground state is, in the dilute limit, very close to the exact ground state. Starting from this ``good" state, the coboson formalism then provides a simple way to reach the exact $N$-coboson ground state.\

 The coboson formalism presented in this work provides a new route for studying scattering properties, Bose-Einstein condensation (BEC), and BEC-BCS cross-over for cold atom gases.

\mbox{}

\section*{Acknowledgments}
This work is supported by the Headquarters of University Advancement at National Cheng-Kung university,  National Science Council of Taiwan under Contract No. NSC 101-2112-M-001-024-MY3, and Academia Sinica, Taiwan. M.C. wishes to thank the National Cheng Kung University and the National Center for Theoretical Sciences (South) for invitations. S.-Y. S. also wishes to thank the Institut des NanoSciences de Paris for invitations.

\end{document}